# Large-scale, Dynamic and Distributed Coalition Formation with Spatial and Temporal Constraints[⋆]

Luca Capezzuto[✉][0000−0003−4404−0998], Danesh Tarapore[0000−0002−3226−6861], and Sarvapali D. Ramchurn[0000−0001−9686−4302]

School of Electronics and Computer Science, University of Southampton
Southampton, UK
{luca.capezzuto,d.s.tarapore,sdr1}@soton.ac.uk

**Abstract** The *Coalition Formation with Spatial and Temporal constraints Problem* (CFSTP) is a multi-agent task allocation problem in which few agents have to perform many tasks, each with its deadline and workload. To maximize the number of completed tasks, the agents need to cooperate by forming, disbanding and reforming coalitions. The original mathematical programming formulation of the CFSTP is difficult to implement, since it is lengthy and based on the problematic Big-M method. In this paper, we propose a compact and easy-to-implement formulation. Moreover, we design D-CTS, a distributed version of the state-of-the-art CFSTP algorithm. Using public London Fire Brigade records, we create a dataset with 347588 tasks and a test framework that simulates the mobilization of firefighters in dynamic environments. In problems with up to 150 agents and 3000 tasks, compared to DSA-SDP, a state-of-the-art distributed algorithm, D-CTS completes $3.79\% \pm [42.22\%, 1.96\%]$ more tasks, and is one order of magnitude more efficient in terms of communication overhead and time complexity. D-CTS sets the first large-scale, dynamic and distributed CFSTP benchmark.

## 1 Introduction

Consider the situation after a disaster, either natural, such as Hurricane Maria in 2017, or man-made, such as the Beirut explosion in 2020. A complex response phase takes place, which includes actions such as extinguishing fires, clearing the streets and evacuating civilians. If the number of first responders is limited, they need to cooperate to act as fast as possible, because any delay can lead to further tragedy and destruction [1]. Cooperation is also necessary when tasks require combined skills. For example, to extract survivors from the rubble of a collapsed building, rescue robots detect life signs with their sensors, firefighters dig and

---

[⋆] Accepted at the 18th European Conference on Multi-Agent Systems (EUMAS 2021).

paramedics load the injured into ambulances. In addition, at any moment new fires could break out or other buildings could collapse, therefore first responders must be ready to deploy to other areas.

Disaster response is a fundamental research topic for multi-agent and multi-robot systems [17,29]. Within this field, we are interested in the *Coalition Formation with Spatial and Temporal constraints Problem* (CFSTP) [5,42]. In the CFSTP, tasks (e.g., save victims or put out fires) have to be assigned to agents (e.g., ambulances or fire brigades). The assignment is determined by the spatial distribution of the tasks in the disaster area, the time needed to reach them, the workload they require (e.g., how large a fire is) and their deadlines (e.g., estimated time left before victims perish). In addition to these constraints, the number of agents may be much smaller than the number of tasks, hence the agents need to cooperate with each other by forming, disbanding and reforming coalitions. A *coalition* is a short-lived and flat organization of agents that performs tasks more effectively or quickly than single agents [5]. The objective of the CFSTP is to define which tasks (e.g., sites with the most victims and the strongest fires) to allocate to which coalitions (e.g., the fastest ambulances and fire trucks with the largest water tanks), in order to complete as many tasks as possible.

Despite having similarities with classic problems such as Generalized Assignment Problem [44] and Job-Shop Scheduling [4], the importance of the CFSTP lies in the fact that it was the first generalization of the Team Orienteering Problem [42, Section 4.2] to consider coalition formation. For this reason, it has been applied in contexts such as human-agent collectives [43], multi-UAV exploration [2] and law enforcement [30].

There are two main issues in the CFSTP literature. First, its original mathematical programming formulation [42, Section 5] is based on 3 sets of binary variables, 1 set of integer variables and 23 types of constraints, 8 of which use the Big-M method. So many variables and constraints make implementation difficult, while the Big-M method introduces a large penalty term that, if not chosen carefully, leads to serious rounding errors and ill conditioning [11]. Second, there is no algorithm that is simultaneously scalable, distributed, and able to solve the CFSTP in systems with a *dynamic environment evolution*[1] (i.e., systems in which, at any time, agents can join in or leave, and new tasks can appear) [10]. Below, we discuss this in detail.

The state-of-the-art CFSTP algorithm, *Cluster-based Task Scheduling* (CTS) [5], transforms the CFSTP into a sequence of $1-1$ task allocations. In other words, instead of allocating each task to a coalition of agents, it forms coalitions by *clustering* or grouping agents based on the closest and most urgent tasks. CTS is anytime (i.e., it returns a partial solution if interrupted before completion), has a polynomial time complexity and can be used in dynamic environments. Its main limitation is being a centralized algorithm. In real-world domains such as disaster response, this leads to three major issues. First, a centralized solver is a single point of failure that makes the system fragile and not robust to unexpected events, such as malfunctions or communication disturbances between agents far apart

---

[1] Also referred to as *open* systems [13]. For brevity, we call them *dynamic environments*.

[35]. Second, if the agents have limited computational resources and the problem is not small, electing a centralized solver might not be possible, while distributing computations always improves scalability. Third, a centralized approach might not be as effective as a distributed approach, given that the situation can evolve rapidly and there could be significant communication delays [27].

To date, only Ramchurn et al. [41] have proposed a dynamic and distributed solution to the CFSTP. They reduced it to a *Dynamic Distributed Constraint Optimization Problem* (DynDCOP) [10] and solved it with *Fast Max-Sum* (FMS), a variant of the Max-Sum algorithm [9] specialized for task allocation. However, unlike CTS, FMS is not guaranteed to convergence, it is not anytime, and its runtime is exponential in the number of agents. Pujol-Gonzalez et al. [37] proposed another Max-Sum variant called *Binary Max-Sum* (BinaryMS), which, compared to FMS, lowers the runtime to polynomial and achieves the same solution quality. Nonetheless, even BinaryMS is not guaranteed to converge and not anytime. In addition, it requires a preprocessing phase with exponential runtime to transform the problem constraints into binary form, which makes it not suitable for dynamic environments. Against this background, we propose the following contributions:

1. A novel mathematical programming formulation of the CFSTP, based only on binary variables and 5 types of constraints, which do not use the Big-M method.
2. D-CTS, a distributed version of CTS that preserves its properties, namely being anytime, scalable and guaranteed to convergence [5].
3. The first large-scale and dynamic CFSTP test framework, based on real-world data published by the London Fire Brigade [22,23].

The rest of the paper is organized as follows. We begin with a discussion of related work in Section 2, then we give our formulation of the CFSTP in Section 3 and present D-CTS in Section 4. Finally, we evaluate D-CTS with our test framework in Section 5 and conclude in Section 6.

## 2 Related work

The CFSTP is NP-hard [42], while CTS is an *incomplete* or non-exact algorithm with a search-based approach [5]. Since we reduce the CFSTP to a DynDCOP in Section 4 and propose a realistic test framework in Section 5, we briefly recall incomplete search-based algorithms and realistic test frameworks, for both DCOPs and DynDCOPs. For a more in-depth look, see [10,19].

### 2.1 Incomplete search-based algorithms

Among the most popular incomplete search-based DCOP algorithms are MGM [24] and DSA [53]. In MGM, each agent iteratively chooses its assignment based on the current neighbor assignments. DSA is an extension of MGM where, to escape from local minima, assignments are chosen stochastically. Both algorithms are efficient, and although they have no quality guarantees on the solutions found, numerous studies have proven their efficacy in many domains. In particular, DSA is a touchstone for novel DCOP algorithms [10]. We use DSA-SDP [54], the state-of-the-art DSA variant, as the baseline in our tests.

Other notable algorithms are $k$-optimal [34], SBDO [3], GDBA [33]. The class of $k$-optimal algorithms decomposes a DCOP into a set of subproblems, each of which involves at most $k$ agents. The solution process continues until no subset of $k$ or fewer agents can improve the global solution. These algorithms are anytime and guaranteed to find a lower bound on the solution quality. However, to eliminate conflicts between partial solutions, each agent may need to communicate with every other agent. Consequently, communication is not local, and both time and space complexity are exponential. Such limitations are also present in the variants proposed in [15,49]. SBDO is a DynDCOP algorithm in which agents exchange arguments about partial solutions. More precisely, each agent tries to send stronger arguments over time to influence its neighbors. Despite being anytime, SBDO has an exponential runtime [10]. GDBA is an extension of the Distributed Breakout Algorithm [51] aimed at solving DCOPs. It is not anytime, but it can be made so by using the Anytime Local Search framework [54]. Moreover, it has polynomial space and time complexity. The results reported in [26,54] suggest that GDBA has similar performance to DSA-SDP.

Dynamic environments pose a challenge to the DCOP research community [19,20,35], to the extent that SBDO and FMS are the only incomplete DynDCOP algorithms proposed to date [10].

### 2.2   Realistic test frameworks

Although the DCOP model can capture numerous real-world problems, researchers usually perform their empirical evaluations on hard random problems or classic combinatorial problems, such as graph coloring and resource allocation [10]. To the best of our knowledge, to date only the following works have conducted tests based on real-world data. Mahesrawan et al. [25] considered resource-constrained multiple-event scheduling problems occurring in office environments. Junges and Bazzan [14] evaluated the performance of complete DCOP algorithms in traffic light synchronization problems. Kim et al. [16] developed heuristics for applying Max-Sum to problems based on the real-time sensor system NetRad. Amador Nelke et al. [30] studied law enforcement problems inspired by police logs. However, none of these test frameworks is as large as ours.

## 3   Problem formulation

We formulate the CFSTP as a *Binary Integer Program* (BIP) [50]. After giving our definitions, we detail our decision variables, constraints and objective function. For constraint programming formulations of the CFSTP, see [5,42].

### 3.1 Definitions

Let $V = \{v_1, \ldots, v_m\}$ be a set of $m$ tasks and $A = \{a_1, \ldots, a_n\}$ be a set of $n$ agents. Let $L$ be the finite set of all possible task and agent locations. Time is denoted by $t \in \mathbb{N}$, starting at $t = 0$, and agents travel or complete tasks with a base time unit of 1. The time units needed by an agent to travel from one location to another are given by the function $\rho : A \times L \times L \to \mathbb{N}$. Having $A$ in the domain of $\rho$ allows to characterize different agent features (e.g., speed or type). Let $l_v$ be the fixed location of task $v$, and let $l_a^t \in L$ be the location of agent $a$ at time $t$, where $l_a^0$ is its initial location and is known a priori.

**Task demand** Each task $v$ has a *demand* $(\gamma_v, w_v)$ such that $\gamma_v$ is the *deadline* of $v$, or the time until which agents can work on $v$ [32], and $w_v \in \mathbb{R}_{\geq 0}$ is the *workload* of $v$, or the amount of work required to complete $v$ [5]. We call $t_{max} = \max_{v \in V} \gamma_v$ the *maximum problem time*.

**Coalition and coalition value** A subset of agents $C \subseteq A$ is called a *coalition*. For each coalition and task there is a *coalition value*, given by the function $u : P(A) \times V \to \mathbb{R}_{\geq 0}$, where $P(A)$ is the power set of $A$. The value of $u(C, v)$ is the amount of work that coalition $C$ does on task $v$ in one time unit. In other words, when $C$ performs $v$, $u(C, v)$ expresses how well the agents in $C$ work together, and the workload $w_v$ decreases by $u(C, v)$ at each time.

### 3.2 Decision variables

Similar to [42, Section 5], we use the following indicator variables:

$$\forall v \in V,\ \forall t \leq \gamma_v,\ \forall C \subseteq A,\ \tau_{v,t,C} \in \{0,1\} \tag{1}$$

$$\forall v \in V,\ \delta_v \in \{0,1\} \tag{2}$$

where: $\tau_{v,t,C} = 1$ if coalition $C$ works on task $v$ at time $t$, and 0 otherwise; $\delta_v = 1$ if task $v$ is completed, and 0 otherwise. Specifying indicator variables for individual agents is not necessary, since they can be inferred from Equation 1.

### 3.3 Constraints

There are 3 types of constraints: structural, temporal and spatial.

**Structural constraints** At each time, at most one coalition can work on each task:

$$\forall v \in V,\ \forall t \leq \gamma_v,\ \sum_{C \subseteq A} \tau_{v,t,C} \leq 1 \tag{3}$$

**Temporal constraints** Tasks can be completed only by their deadlines:

$$\forall v \in V,\ \delta_v \leq 1 \qquad (4)$$

$$\forall v \in V,\ \sum_{t \leq \gamma_v} \sum_{C \subseteq A} u(C, v) \cdot \tau_{v,\,t,\,C} \geq w_v \cdot \delta_v \qquad (5)$$

**Spatial constraints** An agent cannot work on a task before reaching its location. This identifies two cases: when an agent reaches a task from its initial location, and when an agent moves from one task location to another. The first case imposes that, for each task $v$, time $t \leq \gamma_v$ and coalition $C$, the variable $\tau_{v,\,t,\,C}$ can be positive only if all agents in $C$ can reach location $l_v$ at a time $t' < t$:

$$\forall v \in V,\ \forall C \subseteq A,\ \text{if } \lambda = \max_{a \in C} \rho(a, l_a^0, l_v) \leq \gamma_v \text{ then } \sum_{t \leq \lambda} \tau_{v,\,t,\,C} = 0 \qquad (6)$$

$\lambda$ is the maximum time at which an agent $a \in C$ reaches $l_v$, from its initial location at time $t = 0$. Conditional constraints are usually formulated using auxiliary variables or the Big-M method [50]. However, such approaches further enlarge the mathematical program or can cause numerical issues (Section 1). Consequently, in the preprocessing step necessary to create our BIP, we can implement Equation 6 simply by excluding the variables that must be equal to zero.

The second case requires that if an agent cannot work on two tasks consecutively, then it can work on at most one:

$$\forall v_1, v_2 \in V,\ \forall C_1, C_2 \subseteq A \text{ such that } C_1 \cap C_2 \neq \emptyset,$$
$$\forall t_1 \leq \gamma_{v_1},\ \forall t_2 \leq \gamma_{v_2} \text{ such that } t_1 + \max_{a \in C_1 \cap C_2} \rho(a, l_{v_1}, l_{v_2}) \geq t_2, \qquad (7)$$
$$\tau_{v_1,\,t_1,\,C_1} + \tau_{v_2,\,t_2,\,C_2} \leq 1$$

Hence, coalition $C_2$ can work on task $v_2$ only if all agents in $C_1 \cap C_2$ can reach location $l_{v_2}$ by deadline $\gamma_{v_2}$. Equation 7 also implies that an agent cannot work on multiple tasks at the same time.

There are no synchronization constraints [32]. Thus, when a task $v$ is allocated to a coalition $C$, each agent $a \in C$ starts working on $v$ as soon as it reaches its location, without waiting for the remaining agents. This means that $v$ is completed by a temporal sequence of subcoalitions of $C$: $\exists S \subseteq P(C)$ such that $\forall C' \in S,\ \exists t \leq \gamma_v,\ \tau_{v,\,t,\,C'} = 1$, where $P(C)$ is the power set of $C$.

### 3.4 Objective function

Let $\boldsymbol{\tau}$ be a *solution*, that is, a value assignment to all variables, which defines the route and schedule of each agent. The objective is to find a solution that maximizes the number of completed tasks:

$$\arg\max_{\boldsymbol{\tau}} \sum_{v \in V} \delta_v \text{ subject to Equations } 1 - 7 \qquad (8)$$

Both creating all decision variables (Section 3.2) and finding an optimal solution exhaustively (Equation 8) may require to list all $\mathcal{L}$-tuples over $P(A)$, where $\mathcal{L} = |V| \cdot t_{max}$. This implies a worst-case time complexity of:

$$O\left(\left(2^{|A|}\right)^{\mathcal{L}}\right) = O\left(2^{|A| \cdot |V| \cdot t_{max}}\right) \tag{9}$$

**Theorem 1.** *Equation 8 is equivalent to the original mathematical program of the CFSTP [42, Section 5].*

*Proof sketch.* Since we use the original objective function [42, Equation 9], it suffices to verify that our constraints imply the original ones [42, Equations $10 - 32$] as follows. Equations 4 and 5 imply [42, Equations 10 and 11]. Equation 3 implies [42, Equations 12 and 16]. We do not need [42, Equation 13] because $t \leq \gamma_v$ for each $\tau_{v,t,C}$ (Equation 1). Equations 6 and 7, combined with the objective function, imply [42, Equations 14, 15, $17 - 19$]. Equation 7 implies [42, Equations $20 - 22$]. Equations $5 - 7$ imply [42, Equations $25 - 30$]. Equations 3 and 7 imply [42, Equation 31]. Equation 6 implies [42, Equation 32]. □

Having significantly fewer constraints than the original, our BIP can be used more effectively by exact algorithms based on branch-and-cut or branch-and-price [47, Section 3.1.1]. A trivial way to solve the CFSTP would be to implement Equation 8 with solvers such as CPLEX or GLPK. Although this would guarantee anytime and optimal solutions, it would also take exponential time to both create and solve our BIP (Equation 9). This limits this practice to offline contexts or very small problems. For example, using CPLEX 20.1 with commodity hardware and the test setup of [42], we can solve problems where $|A| \cdot |V| \leq 50$ in hours. With bigger problems, the runtime increases rapidly to days.

Another major issue with centralized generation of optimal solutions is that, in real-time domains such as disaster response, it can be computationally not feasible (Section 1) or economically undesirable, especially when the problem changes frequently [5]. For these reasons, the next section presents a scalable, dynamic and distributed algorithm.

## 4 A scalable, dynamic and distributed CFSTP algorithm

We reduce the CFSTP to a DynDCOP, then we show how CTS, the state-of-the-art CFSTP algorithm [5], can solve it. We use the DynDCOP formalism because it has proven largely capable of modeling disaster response problems [10].

### 4.1 Reduction of the CFSTP to a DynDCOP

Following [10], we formalize a DynDCOP as a sequence $\mathcal{D} = \{\mathcal{D}_t\}_{t \leq t_{max}}$, where each $\mathcal{D}_t = (A^t, X^t, D^t, F^t)$ is a DCOP such that $A^t \subseteq A$ and:

- $X^t = \{x_1^t, \ldots, x_k^t\}$ is a set of $k = |A^t| \leq n$ variables, where $x_i^t$ is the task performed by agent $a_i^t \in A^t$.

- $D^t = \{D_1^t, \ldots, D_k^t\}$ is a set of $k$ variable domains, such that $x_i^t \in D_i^t$. A set $d = \{d_1, \ldots, d_k\}$, where $d_i \in D_i^t$, is called an *assignment*. Each $d_i \in d$ is called the $i$-th *variable assignment* and is the value assigned to variable $x_i^t$.
- $F^t = \{f_1^t, \ldots, f_h^t\}$ is a set of $h \leq m$ functions, where $f_i^t$ represents the constraints on task $v_i^t$. In particular, each $f_i^t : D_{i_1}^t \times \cdots \times D_{i_{h_i}}^t \to \mathbb{R}_{\geq 0}$ assigns a non-negative real cost to each possible assignment to the variables $X_{h_i}^t \subseteq X^t$, where $h_i \leq h$ is the arity of $f_i^t$.

The objective is to find an assignment that minimizes all costs:

$$\forall t \leq t_{max}, \arg\min_{d \in D^t} \sum_{f_i^t \in F^t} f_i^t(d_{i_1}, \ldots, d_{i_{h_i}}) \tag{10}$$

It is typically assumed that if $x_i^t$ is in the scope of $f_j^t$, then agent $a_i^t$ knows $f_j^t$ [10, Section 4.2]. To reduce the CFSTP to a DynDCOP, we define $A^t$, $D^t$ and $F^t$ as follows. At time $t$, let $A^t$ be the set of agents that are not working on nor traveling to a task (i.e., *free* or idle agents [5]), and let $V_{allocable}^t$ be the set of tasks that have not yet been completed. The domain of each variable $x_i^t$ is:

$$D_i^t = \left\{ v \in V_{allocable}^t \text{ such that } t + \rho(a_i^t, l_{a_i^t}, l_v) \leq \gamma_v \right\} \cup \{\varnothing\} \tag{11}$$

where $\varnothing$ means that no task is allocated to agent $a_i^t$. Hence, $A^t$ satisfies the structural constraints, while $D_i^t$ contains the tasks that at time $t$ can be allocated to $a_i^t$ satisfying the spatial constraints (Section 3.3). Let $\boldsymbol{\tau}_i \subseteq \boldsymbol{\tau}$ be a *singleton solution*, that is, a solution to task $v_i$. At time $t$, let $\boldsymbol{\tau}_i^t \subseteq \boldsymbol{\tau}_i$ be a singleton solution corresponding to $f_i^t(d_{i_1}, \ldots, d_{i_{h_i}})$, defined as follows. Each $\tau_{v_i, t, C} \in \boldsymbol{\tau}_i^t$ is such that $C$ is a subset of the agents that control the variables in the scope of $f_i^t$, while $\tau_{v_i, t, C} = 1$ if $d_{i_{h_i}} = v_i$, for each $h_i$-th agent in $C$, and 0 otherwise. To satisfy the temporal constraints (Section 3.3), each $i$-th function is defined as follows:

$$f_i^t(d_{i_1}, \ldots, d_{i_{h_i}}) = \min_{\boldsymbol{\tau}_i^t, t' \leq \gamma_{v_i}} \sum_{s \leq t', \tau_{v_i, s, C} \in \boldsymbol{\tau}_i^t} u(C, v) \geq w_v \tag{12}$$

with the convention that $f_i^t(d_{i_1}, \ldots, d_{i_{h_i}}) = +\infty$ if $v_i$ cannot be completed by deadline $\gamma_v$. Hence, the solution space of $\mathcal{D}$ satisfies all CFSTP constraints, while minimizing all costs implies minimizing the time required to complete each task (Equations 10 and 12), which implies maximizing the total number of completed tasks, as required by the objective function of the CFSTP (Equation 8).

### 4.2  Distributed CTS

At each time, CTS executes in sequence the following two phases [5]:

1. For each free agent $a$, associate $a$ with an uncompleted task $v$ such that $v$ is the closest to $a$ and deadline $\gamma_v$ is minimum.
2. For each uncompleted task $v$, allocate $v$ to a coalition $C$ such that $|C|$ is minimum and each agent $a \in C$ has been associated with $v$ in Phase 1.

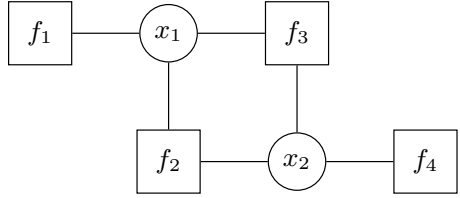

**Figure 1.** The factor graph of a DCOP with 2 agents and 4 tasks. In our formulation, a DCOP represents the state of a CFSTP at a certain time, in which circles are variables of free agents, squares are cost functions of uncompleted tasks, and each edge connects an agent to a task it can reach by its deadline.

To represent a DCOP, we use a *factor graph* [18,21], which decomposes the problem into three parts: *variable nodes*, representing the variables; *factor nodes*, representing the constraints; undirected edges between each factor node and the variable nodes in its scope. As an example, Figure 1 shows the factor graph of the function $F(X) = f_1(x_1) + f_2(x_1, x_2) + f_3(x_1, x_2) + f_4(x_2)$.

In a factor graph $G$, a solution is found by allowing nodes to exchange messages. Hence, to execute CTS on $G$, we have to define how the nodes communicate and operate. Below, we present a communication protocol and algorithms for both variable and factor nodes. Based on the well-established formalism of Yokoo et al. [52], the nodes communicate in the following way:

- Node $i$ can message node $j$ only if $i$ knows the address of $j$. In our context, if $x_i^t$ is in the scope of $f_j^t$, then $x_i^t$ knows the address of $f_j^t$, and vice versa.
- Each node $i$ has a message queue $Q_i$, to which messages are delivered with a finite delay.
- Node $i$ can use the function RECEIVE() to dequeue a message from $Q_i$, and the function SEND($j$, `illoc_force`, `[args]`) to send a message to $j$. Node $j$ will receive a message in the format (`sender`, `illoc_force`, `[args]`), where `sender` is the identifier of node $i$, `illoc_force` is its illocutionary force, and `[args]` is an optional list of arguments. By *illocutionary force*, we mean either an information or a command [48].

We assume that the node of each function is controlled by an agent in its scope. Algorithm 1 presents the operation of variable node $x_i^t$. If there is an uncompleted task $v_j^t$ that can be allocated to free agent $a_i^t$ (lines $1 - 3$), then variable node $x_i^t$ communicates to factor node $f_j^t$ the ability of $a_i^t$ to work on $v_j^t$, also specifying the time at which it can reach and start working on it (lines $4 - 6$). After that, it waits until it gets a reply from $f_j^t$ or a predetermined time interval expires (lines $7 - 9$). If it receives the approval of $f_j^t$, then $v_j^t$ is allocated to $a_i^t$ (lines $10 - 11$). At line 2, $v_j^t$ is chosen such that it is the closest to $a_i^t$ and $\gamma_{v_j^t}$ is the shortest deadline [5]. Phase 1 is completed after that each $x_i^t$ executes line 6.

Algorithm 2 presents the operation of factor node $f_j^t$. The loop at lines $1 - 2$ is a synchronization step that allows $f_j^t$ to know which agents in its neighborhood can work on $v_j^t$. Lines $3 - 6$ enacts Phase 2, while lines $7 - 9$ update workload $w_{v_j}$.

---

**Algorithm 1:** CTS node of variable $x_i^t$

---

1  $x_i^t \leftarrow \varnothing$ ▷ initialize to *idle*
2  $d_j \leftarrow$ get task allocable to agent $a_i^t$ at time $t$ ▷ [5, Algorithm 5]
3  **if** $d_j \neq \varnothing$ **then**
4  | $s_i \leftarrow$ time at which agent $a_i^t$ can start working on task $d_j$
5  | $f_j^t \leftarrow$ factor node of $d_j$
6  | SEND($f_j^t$, assignable, $s_i$)
7  | msg $\leftarrow$ NIL
8  | **while** *msg not received from $f_j^t$ or not time out* **do**
9  | | msg $\leftarrow$ RECEIVE()
10 | **if** *msg* $= (f_j^t,$ *allocate*$)$ **then**
11 | | $x_i^t \leftarrow d_j$

---

**Algorithm 2:** CTS node of factor $f_j^t$

---

1  **while** *not all neighbors sent an* **assignable** *message or not time out* **do**
2  | msg $\leftarrow$ RECEIVE()
3  $\Pi_{v_j}^t \leftarrow$ list of all assignable agents sorted by arrival time to $v_j$
4  $C^* \leftarrow$ minimum coalition in $\Pi_{v_j}^t$ that can complete $v_j$ by $\gamma_v$ ▷ Equation 12
5  **for** $a_i^t \in C^*$ **do**
6  | SEND($x_i^t$, allocate)
7  $C_{v_j}^t \leftarrow$ all agents working on $v_j$ at time $t$
8  **if** $C_{v_j}^t \neq \emptyset$ **then**
9  | $w_{v_j} \leftarrow w_{v_j} - u(C_{v_j}^t, v_j)$

---

We call *Distributed CTS* (D-CTS) the union of Algorithms 1 and 2. The size of each message is $O(1)$, since it always contains a node address, a message flag and an integer. At time $t$, each variable node $x_i^t$ sends at most 1 message (line 6 in Algorithm 1), while each factor node $f_j^t$ sends $O(|A|)$ messages (lines $5-6$ in Algorithm 2). Assuming that all tasks can be completed, the total number of messages sent is $O(|A| + |V| \cdot |A|) = O(|V| \cdot |A|)$.

The runtime of Algorithm 1 is $O(|V|)$, because line 2 selects a task in the neighborhood of an agent. The runtime of Algorithm 2 is $O(|A| \log |A|)$, due to the sorting at line 3 [8]. Since both algorithms are executed up to $t_{max}$ times, the overall time complexity of D-CTS is the same as CTS [5, Equations 10 and 11]:

$$\Omega\left(t_{max} \cdot (|V| + |A| \log |A|)\right) \text{ and } O\left(t_{max} \cdot |V| \cdot |A| \log |A|\right) \qquad (13)$$

where the lower bound represents the case in which the operations of each phase are executed in parallel. The advantages of D-CTS are as follows:

1. It is anytime, since it decomposes a CFSTP into a set of independent subproblems (Section 1). This property is not trivial to guarantee in distributed

    systems [54], and is missing in main DCOP algorithms (e.g., ADOPT, DPOP, OptAPO and Max-Sum [10, Table 4]).
2. It is self-stabilizing [10, Definition 6], being guaranteed to converge [5, Theorem 1], and given that each agent can only work on a new task after completing the one to which it is currently assigned (Algorithm 1).
3. The phase-based design has two performance benefits. First, the algorithm is not affected by the structure of factor graphs. For instance, in a cyclic graph like the one in Figure 1, where the same $a > 1$ tasks can be allocated to the same $b > 1$ agents, inference-based DCOP algorithms (e.g., Max-Sum and BinaryMS) in general are not guaranteed to converge, unless they are augmented with specific techniques (e.g., damping [7] or ADVP [55]). Second, the algorithm is robust to *disruptions*, that is, to the addition or removal of nodes from a factor graph [41, Section 6.2]. Disruptions are typical of real-world domains [5]. For instance, in disaster response, tasks are removed if some victims have perished, and are added if new fires are discovered. Likewise, new agents are added to reflect the availability of additional workforce, while existing ones are removed when they deplete their resources or are unable to continue due to sustained damages. Unlike D-CTS, the majority of DCOP algorithms (e.g., Max-Sum and DPOP) cannot handle disruptions, unless they are properly modified or extended (e.g., FMS and S-DPOP [10]). Hence, besides being a DynDCOP algorithm, D-CTS can also cope with runtime changes in a DCOP formulation.
4. Unlike most DCOP algorithms (e.g., ADOPT and DPOP), the communication overhead (i.e., the number of messages exchanged) is at most linear, and each agent does not need to maintain an information graph of all other agents.
5. Finally, performance does not depend on any tuning parameters, as is the case with other algorithms (e.g., DSA variants).

## 5 Empirical evaluation in dynamic environments

We created a dataset[2] with 347588 tasks using open records published by the London Fire Brigade over a period of 11 years. Then, we wrote a test framework in Java[3] and compared D-CTS against DSA-SDP [54], a state-of-the-art incomplete, synchronous and search-based DCOP algorithm.

    We adapted DSA-SDP to solve our DynDCOP formulation (Section 4.1), which decomposes the CFSTP into a sequence of independent subproblems. Hence, although originally a DCOP algorithm, its performance is not penalized in our test framework. We chose it as our baseline because, similarly to D-CTS, it has a polynomial coordination overhead and is scalable (Section 2). We kept the parameters of [54] and ran $|V^t_{allocable}|$ iterations at each time $t$, since we found that, in our test framework, running more iterations can only marginally improve the solution quality, while requiring a significant increase in communication overhead and time complexity. Below, we detail our setup and discuss the results.

---

[2] https://zenodo.org/record/4728012
[3] https://zenodo.org/record/4764646

### 5.1 Setup

Let $\mathcal{N}$ and $\mathcal{U}$ denote the normal and uniform distribution, respectively. A test configuration consists of the following parameters:

- Since there are currently 150 identical London fire engines in operation, $|A| = 150$ for each problem. All agents have the same speed, but each may perform differently in different coalitions.
- $|V| = |A| \cdot k$, where $k \in \mathbb{N}^+$ and $k \leq 20$. Thus, problems have up to 3000 tasks.
- Each task $v$ is a fire or a special service, and its demand is defined by a record dated between 1 January 2009 and 31 December 2020. More precisely, $\gamma_v$ is the attendance time (in seconds) of the firefighters, and since the median attendance time in the whole dataset is about 5 minutes, we set $w_v \sim \mathcal{U}(10, 300)$ to simulate wide ranging workloads.
- For each task-to-agent ratio $|V|/|A|$, the nodes of a problem are chosen in chronological order. That is, the first problem always starts with record 1, and if a problem stops at record $q$, then the following one will use records $q+1$ to $q+1+|V|$.
- The locations are latitude-longitude points, and the travel time $\rho(a, l_1, l_2)$ is given by the distance between locations $l_1$ and $l_2$ divided by the (fixed) speed of agent $a$.
- In addition to task locations, $L$ contains the locations of the 103 currently active London fire stations. In each problem, each agent starts at a fire station defined by the record of a task.
- To generate coalition values, we start by taking from [38, Section 4] the following well-known distributions:
  1. *Normally Distributed Coalition Structures* (NDCS): $u(C, v) \sim \mathcal{N}(|C|, \sqrt[4]{|C|})$.
  2. *Agent-based*: each agent $a$ has a value $p_a \sim \mathcal{U}(0, 10)$ representing its individual performance and a value $p_a^C \sim \mathcal{U}(0, 2 \cdot p_a)$ representing its performance in coalition $C$. The value of a coalition is the sum of the values of its members: $u(C, v) = \sum_{a \in C} p_a^C$.

  Then, we decrease each $\mu_v = u(C, v)$ by $r \sim \mathcal{U}(\mu_v/10, \mu_v/4)$ with probability $\gamma_v/(t_{max}+1)$, and by $q \sim \mathcal{U}(\mu_v/10, \mu_v/4)$ with probability $|C|/(|A|+1)$. The perturbation $r$ simulates real-time domains, where the earlier the deadline for a task, the higher the reward [45]. The perturbation $q$ simulates situations where the more agents there are, the greater the likelihood of congestion and thus of reduced performance, as it can happen in large-scale robot swarms [12]. We call the resulting distributions UC_NDCS and UC_Agent-based, where UC means *Urgent and Congested*. NDCS does not to favor solutions containing fewer coalitions [40], while Agent-based tends to do the opposite. By using them, we obtain solution spaces in which higher values are first associated with smaller coalitions and then with larger coalitions. Both distributions are neither superadditive nor subadditive [39]. Hence, it is not possible to define a priori an optimal coalition for each task.

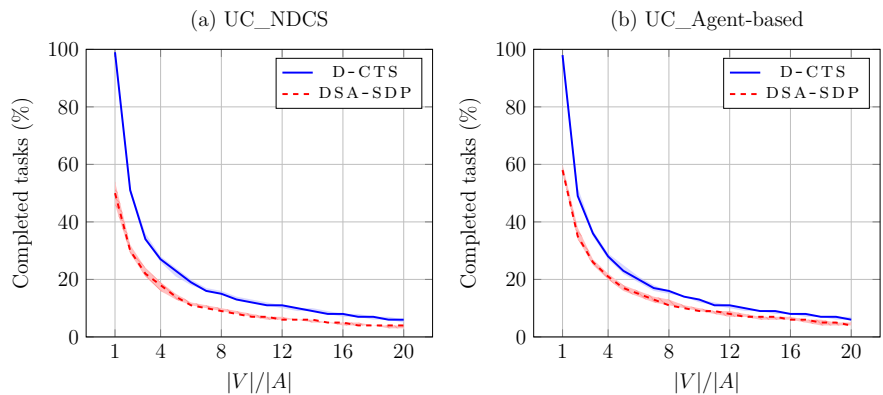

**Figure 2.** Performance of DSA-SDP and D-CTS in our test framework. Each subfigure denotes a coalition value distribution, while each point is the median and 95% confidence interval over 100 problems of the percentage of tasks completed. The X-axis is the task-to-agent ratio.

During the solution of each problem, we gradually removed agents to simulate *degradation* scenarios. The removal rate was calculated with a Poisson cumulative distribution function $Pois_{CDF}(\boldsymbol{a}, \lambda)$, where $\boldsymbol{a}$ contains all firefighter arrival times in the dataset, and the rate $\lambda$ is the average number of incidents per hour and per day. For each test configuration and algorithm, we solved 100 problems and measured the median and 95% confidence interval of: number of messages sent; *network load*, or the total size of messages sent; number of *Non-Concurrent Constraint Checks* (NCCCs) [28]; percentage of tasks completed, and CPU time[4].

### 5.2  Results

Figure 2 and 3 show our results. D-CTS completes $3.79\% \pm [42.22\%, 1.96\%]$ more tasks than DSA-SDP (Figure 2). For both algorithms, the performance drops rapidly as the task-to-agent ratio increases. This is due to the Urgent component in the coalition value distributions: the higher the ratio, the higher the median task completion time. Conversely, the Congested component can reduce the percentage of tasks completed more in problems with smaller task-to-agent ratios, where agents can form larger coalitions and thus increase the likelihood of congestion.

The network load of DSA-SDP is $0.59 \pm [0.41, 0.02]$ times that of D-CTS (Figure 3b). This is because a DSA-SDP message contains only a task address, while a D-CTS message also contains a binary flag and an integer (Section 4.2). In Java, an address requires 8 bytes, a flag requires 1 byte, and an integer requires $1 - 4$ bytes. Hence, while a DSA-SDP message always requires 8 bytes, a D-CTS message requires $10 - 13$ bytes. This is line with the results obtained. However, the situation would be reversed if we performed 1000 DSA-SDP iterations as suggested in [53], since $median(\{|V_{allocable}^t|\}_{t \leq t_{max}}) \ll 1000$ in our tests.

---

[4] Based on an Intel Xeon E5-2670 processor (octa-core 2.6 GHz with Hyper-Threading).

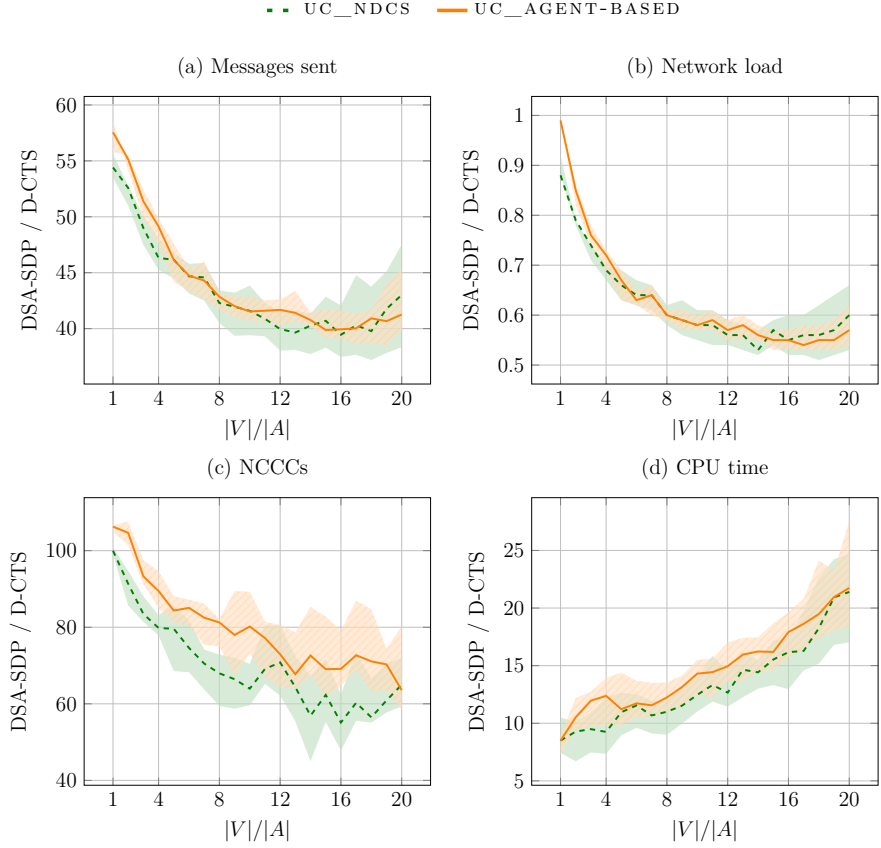

**Figure 3.** Ratio of DSA-SDP performance to D-CTS performance. Each subfigure denotes a performance metric $m$, while each point is the median and 95% confidence interval over 100 problems of $m_a/m_b$, where $m_a$ (resp. $m_b$) is the value of DSA-SDP (resp. D-CTS) for $m$. The X-axis is the task-to-agent ratio.

The remaining metrics put DSA-SDP at a distinct disadvantage (Figure 3a, c, d). The overload compared to D-CTS is $41.72 \pm [12.45, 0.42]$ times more messages sent, $72.78 \pm [34.79, 27.79]$ times more NCCCs, and $13.82 \pm [4.52, 3.71]$ times more CPU time. This is explained as follows. While the number of messages sent is $O(|V| \cdot |A|)$ in D-CTS (Section 4.2), it is $O(|V| \cdot |A|^2)$ in DSA-SDP, since the agents exchange their assignments [54]. In D-CTS, analyzing in sequence the agents that can be assigned to each task (line 4 in Algorithm 2) requires $O(|V| \cdot |A|)$ NCCCs. DSA-SDP does a similar analysis, but for each message exchanged between two agents, which requires $O(|V|^2 \cdot |A|^2)$ NCCCs. Finally, the time complexity of DSA-SDP is $O(t_{max} \cdot |V| \cdot |A|^2)$, where $O(|V| \cdot |A|)$ is required by the message

exchange phase at each time, and $O(|A|)$ is required by each agent to calculate the assignment costs (Equation 12). Hence, DSA-SDP is asymptotically slower than D-CTS (Equation 13). Overall, D-CTS took $525 \pm [281, 482]$ ms, while DSA-SDP took $6.97 \pm [5.84, 6.2]$ seconds. In accordance with the above, the ratio of DSA-SDP performance to D-CTS performance tends to increase with regard to CPU time, and to decrease with regard to the other metrics.

In a dynamic environment, desirable features of a distributed algorithm include being robust to disruptions and minimizing communication overhead (Section 4.2). The latter feature is particularly important in real-world domains such as disaster response, where agent communication can be costly (i.e., non *free-comm* environment [36]) or there might be operational constraints, such as low bandwidth or limited network topology (e.g., sparse robot swarms searching for shipwrecks on the seabed or monitoring forest fires [46]). In our tests, compared to DSA-SDP, D-CTS achieves a slightly better solution quality (Figure 2), and is one order of magnitude more efficient in terms of communication overhead and time complexity (Figure 3). This affirms its effectiveness as a scalable and distributed CFSTP algorithm for dynamic environments.

## 6   Conclusions

We gave a novel mathematical programming formulation of the CFSTP, which is significantly shorter and easier to implement than the original [42]. By reducing the CFSTP to a DynDCOP, we also designed D-CTS, the first distributed version of the state-of-the-art CFSTP algorithm. Finally, using real-world data provided by the London Fire Brigade and a large-scale test framework, we compared D-CTS against DSA-SDP, a state-of-the-art distributed algorithm. In situations where the number of agents monotonically decreases over time, D-CTS has slightly better median performance, as well as significantly lower communication overhead and time complexity. Future work aims at extending our test framework by:

1. Comparing D-CTS with other state-of-the-art distributed algorithms, such as DALO [15], SBDO [3], GDBA [33], D-Gibbs [31] and FMC_TA [30].
2. Adding more realistic coalition value distributions.
3. Studying *exploration* scenarios [10], that is, designing tests in which tasks are gradually added to the system.

We also want to transfer our work to the MARSC model [6], which, unlike the CFSTP, can capture situations where there are soft deadlines, tasks are not all equally important, and there may be an order of completion.

Finally, given its advantages (Section 4.2) and the scarcity of incomplete DynDCOP algorithms (Section 2), we want to design a D-CTS extension with provable bounds on solution quality and able to solve general DynDCOPs.

## Acknowledgments

We thank Alessandro Farinelli for his suggestions, and the anonymous reviewers for helping us correct and improve the manuscript. This research is sponsored by UKRI and AXA Research Fund. Luca Capezzuto acknowledges the use of the IRIDIS High Performance Computing facility at the University of Southampton.

## References

1. Alexander, D.E.: Principles of Emergency Planning and Management. Oxford University Press (2002)
2. Baker, C.A.B., Ramchurn, S., Teacy, W.L., Jennings, N.R.: Planning search and rescue missions for uav teams. In: ECAI. pp. 1777–1782 (2016)
3. Billiau, G., Chang, C.F., Ghose, A.: Sbdo: A new robust approach to dynamic distributed constraint optimisation. In: Int. Conf. on Princ. and Pract. of MAS. pp. 11–26. Springer Berlin Heidelberg (2012)
4. Brucker, P.: Scheduling Algorithms. Springer-Verlag, 5th edn. (2007)
5. Capezzuto, L., Tarapore, D., Ramchurn, S.D.: Anytime and efficient multi-agent coordination for disaster response. SN Comput. Sci. **2**(165) (2021)
6. Capezzuto, L., Tarapore, D., Ramchurn, S.D.: Multi-agent routing and scheduling through coalition formation. In: OptLearnMAS-21 (2021)
7. Cohen, L., Zivan, R.: Max-sum revisited: The real power of damping. In: AAMAS. pp. 111–124. Springer (2017)
8. Cormen, T.H., Leiserson, C.E., Rivest, R.L., Stein, C.: Introduction to Algorithms. MIT press, 3rd edn. (2009)
9. Farinelli, A., Rogers, A., Petcu, A., Jennings, N.R.: Decentralised coordination of low-power embedded devices using the max-sum algorithm. In: AAMAS. vol. 2, pp. 639–646 (2008)
10. Fioretto, F., Pontelli, E., Yeoh, W.: Distributed constraint optimization problems and applications: A survey. JAIR **61**, 623–698 (2018)
11. Griva, I., Nash, S.G., Sofer, A.: Linear and Nonlinear Optimization. Society for Industrial and Applied Mathematics, 2nd edn. (2009)
12. Guerrero, J., Oliver, G., Valero, O.: Multi-robot coalitions formation with deadlines: Complexity analysis and solutions. PloS one **12**(1) (2017)
13. Hewitt, C.: The challenge of open systems, pp. 383–395. Cambridge University Press (1990)
14. Junges, R., Bazzan, A.L.C.: Evaluating the performance of dcop algorithms in a real world, dynamic problem. In: AAMAS. vol. 2, pp. 599–606 (2008)
15. Kiekintveld, C., Yin, Z., Kumar, A., Tambe, M.: Asynchronous algorithms for approximate distributed constraint optimization with quality bounds. In: AAMAS. p. 133–140 (2010)
16. Kim, Y., Krainin, M., Lesser, V.: Effective variants of the max-sum algorithm for radar coordination and scheduling. In: IEEE/WIC/ACM Int. Confs. on Web Int. and Int. Ag. Tech. vol. 2, pp. 357–364 (2011)
17. Kitano, H., Tadokoro, S.: Robocup rescue: A grand challenge for multiagent and intelligent systems. AI Mag. **22**(1), 39–39 (2001), https://rescuesim.robocup.org
18. Kschischang, F.R., Frey, B.J., Loeliger, H.A.: Factor graphs and the sum-product algorithm. IEEE Trans. on Inf. Th. **47**(2), 498–519 (2001)
19. Leite, A.R., Enembreck, F., Barthes, J.P.A.: Distributed constraint optimization problems: Review and perspectives. Exp. Sys. w. App. **41**(11), 5139–5157 (2014)
20. Lesser, V., Corkill, D.: Challenges for multi-agent coordination theory based on empirical observations. In: AAMAS. p. 1157–1160 (2014)

21. Loeliger, H.A.: An introduction to factor graphs. IEEE Sig. Proc. Mag. **21**(1), 28–41 (2004)
22. London Datastore: London Fire Brigade Incident Records (2021), https://data.london.gov.uk/dataset/london-fire-brigade-incident-records
23. London Datastore: London Fire Brigade Mobilisation Records (2021), https://data.london.gov.uk/dataset/london-fire-brigade-mobilisation-records
24. Maheswaran, R.T., Pearce, J.P., Tambe, M.: Distributed algorithms for dcop: A graphical-game-based approach. In: Int. Conf. on Par. and Distr. Comp. Sys. pp. 432–439 (2004)
25. Maheswaran, R.T., Tambe, M., Bowring, E., Pearce, J.P., Varakantham, P.: Taking dcop to the real world: Efficient complete solutions for distributed multi-event scheduling. In: AAMAS. vol. 1, pp. 310—317 (2004)
26. Mahmud, S., Khan, M.M., Jennings, N.R.: On population-based algorithms for distributed constraint optimization problems. arXiv:2009.01625 (2020)
27. Mailler, R., Zheng, H., Ridgway, A.: Dynamic, distributed constraint solving and thermodynamic theory. JAAMAS **32**(1), 188–217 (2018)
28. Meisels, A.: Distributed Search by Constrained Agents. Springer (2007)
29. Murphy, R.R.: Disaster Robotics. MIT press (2014)
30. Nelke, S.A., Okamoto, S., Zivan, R.: Market clearing-based dynamic multi-agent task allocation. ACM Trans. on Int. Sys. and Tech. **11**(1), 1–25 (2020)
31. Nguyen, D.T., Yeoh, W., Lau, H.C., Zivan, R.: Distributed gibbs: A linear-space sampling-based dcop algorithm. JAIR **64**, 705–748 (2019)
32. Nunes, E., Manner, M., Mitiche, H., Gini, M.: A taxonomy for task allocation problems with temporal and ordering constraints. JRAS **90**, 55–70 (2017)
33. Okamoto, S., Zivan, R., Nahon, A., et al.: Distributed breakout: Beyond satisfaction. In: IJCAI. pp. 447–453 (2016)
34. Pearce, J.P., Tambe, M.: Quality guarantees on k-optimal solutions for distributed constraint optimization problems. In: IJCAI. pp. 1446–1451 (2007)
35. Petcu, A.: A class of algorithms for distributed constraint optimization. Ph.D. thesis, École polytechnique fédérale de Lausanne (2007)
36. Ponda, S.S., Johnson, L.B., Geramifard, A., How, J.P.: Cooperative Mission Planning for Multi-UAV Teams. Handbook of Unmanned Aerial Vehicles, pp. 1447–1490. Springer (2015)
37. Pujol-Gonzalez, M., Cerquides, J., Farinelli, A., Meseguer, P., Rodriguez-Aguilar, J.A.: Efficient inter-team task allocation in robocup rescue. In: AAMAS. pp. 413–421 (2015)
38. Rahwan, T., Michalak, T., Jennings, N.: A hybrid algorithm for coalition structure generation. In: AAAI. vol. 26 (2012)
39. Rahwan, T., Michalak, T.P., Wooldridge, M., Jennings, N.R.: Coalition structure generation: A survey. AI **229**, 139–174 (2015)
40. Rahwan, T., Ramchurn, S.D., Jennings, N.R., Giovannucci, A.: An anytime algorithm for optimal coalition structure generation. JAIR **34**, 521–567 (2009)
41. Ramchurn, S.D., Farinelli, A., Macarthur, K.S., Jennings, N.R.: Decentralized coordination in robocup rescue. The Computer Journal **53**(9), 1447–1461 (2010)
42. Ramchurn, S.D., Polukarov, M., Farinelli, A., Truong, C., Jennings, N.R.: Coalition formation with spatial and temporal constraints. In: AAMAS. pp. 1181–1188 (2010)
43. Ramchurn, S.D., Wu, F., Jiang, W., Fischer, J.E., Reece, S., Roberts, S., Rodden, T., Greenhalgh, C., Jennings, N.R.: Human-agent collaboration for disaster response. JAAMAS **30**(1), 82–111 (2016)

44. Ross, G.T., Soland, R.M.: A branch and bound algorithm for the generalized assignment problem. Math. Prog. **8**(1), 91–103 (1975)
45. Stankovic, J.A., Spuri, M., Ramamritham, K., Buttazzo, G.C.: Deadline scheduling for real-time systems: EDF and related algorithms, vol. 460. Springer Science & Business Media (2013), reprint of the original 1998 edition
46. Tarapore, D., Groß, R., Zauner, K.P.: Sparse robot swarms: Moving swarms to real-world applications. Front. in Rob. and AI **7**,  83 (2020)
47. Vansteenwegen, P., Gunawan, A.: Orienteering Problems: Models and Algorithms for Vehicle Routing Problems with Profits. Springer Nature, Switzerland (2019)
48. Vieira, R., Moreira, Á.F., Wooldridge, M., Bordini, R.H.: On the formal semantics of speech-act based communication in an agent-oriented programming language. JAIR **29**, 221–267 (2007)
49. Vinyals, M., Shieh, E., Cerquides, J., Rodriguez-Aguilar, J.A., Yin, Z., Tambe, M., Bowring, E.: Quality guarantees for region optimal dcop algorithms. In: AAMAS. vol. 1, p. 133–140 (2011)
50. Wolsey, L.A.: Integer Programming. John Wiley & Sons, second edn. (2020)
51. Yokoo, M., Hirayama, K.: Distributed breakout algorithm for solving distributed constraint satisfaction problems. In: Proc. of the 2nd Int. Conf. on MAS. pp. 401–408. MIT Press Cambridge (1996)
52. Yokoo, M., Ishida, T., Durfee, E.H., Kuwabara, K.: Distributed constraint satisfaction for formalizing distributed problem solving. In: Proc. of the 12th Int. Conf. on Distr. Comp. Sys. pp. 614–621. IEEE (1992)
53. Zhang, W., Wang, G., Xing, Z., Wittenburg, L.: Distributed stochastic search and distributed breakout: properties, comparison and applications to constraint optimization problems in sensor networks. AI **161**(1-2), 55–87 (2005)
54. Zivan, R., Okamoto, S., Peled, H.: Explorative anytime local search for distributed constraint optimization. AI **212**, 1–26 (2014)
55. Zivan, R., Parash, T., Cohen, L., Peled, H., Okamoto, S.: Balancing exploration and exploitation in incomplete min/max-sum inference for distributed constraint optimization. JAAMAS **31**(5), 1165–1207 (2017)